\documentstyle[sprocl,epsfig]{article}

\def\be{\begin{equation}}
\def\ee{\end{equation}}
\def\bea{\begin{eqnarray}}
\def\eea{\end{eqnarray}}

\begin{document}

\title{Recent Status of Neutrino Oscillation Study}

\author{OSAMU YASUDA}

\address{Department of Physics,
Tokyo Metropolitan University \\
Minami-Osawa, Hachioji, Tokyo 192-0397, Japan
\\E-mail: yasuda@phys.metro-u.ac.jp}


\maketitle\abstracts{
I review briefly the recent status of research on neutrino
oscillations, such as three and four flavor analysis of
atmospheric neutrinos and solar neutrinos, and prospects of present and
future long baseline experiments.}

\section{Introduction}

It has been known that the atmospheric neutrino anomaly
(See, e.g., Ref.~\cite{Kajita:2001mr})
can be accounted for by dominant $\nu_\mu\leftrightarrow\nu_\tau$
oscillations with almost maximal mixing.
According to the most up-to-date result of the two flavor
analysis of $\nu_\mu\leftrightarrow\nu_\tau$ with 1289 day
SuperKamiokande data \cite{Toshito:2001dk}, the allowed region
of the oscillation parameters at 90\%CL is
$0.88<\sin^22\theta_{\mbox{\rm\footnotesize atm}}\le 1$,
$1.6\times10^{-3}{\rm eV}^2<
\Delta m_{\mbox{\rm\footnotesize atm}}^2 < 4\times10^{-3}{\rm
eV}^2$.
Two flavor analysis of $\nu_\mu\leftrightarrow\nu_s$ has been also done
by the SuperKamiokande group using the data of
neutral current enriched multi-ring events,
high energy partially contained events and upward going $\mu$'s,
and they have excluded the two flavor oscillation
$\nu_\mu\leftrightarrow\nu_s$ at 99\%CL \cite{SKs}.

It has been also shown that the solar neutrino
observations (See, e.g., Ref.~\cite{Bahcall:2001nu})
suggest neutrino oscillations and the most up-to-date analyses
\cite{Fogli:2001vr,Bahcall:2001zu,Bandyopadhyay:2001aa,Krastev:2001tv}
tell us that
the large mixing angle (LMA) MSW solution
($\Delta m^2_\odot\simeq 2\times 10^{-5}$eV$^2$, $\sin^22\theta\simeq 0.8$)
gives the best fit and it is followed by the LOW solution
($\Delta m^2_\odot\simeq 1\times 10^{-7}$eV$^2$, $\sin^22\theta\simeq 1.0$).
The recent SNO data \cite{Ahmad:2001an}
prefer $\nu_e\leftrightarrow\nu_{\rm active}$
to $\nu_e\leftrightarrow\nu_s$ oscillations.

On the other hand, it has been claimed by the LSND group that their data \cite{Mills:2001tq}
suggest neutrino oscillations with
$\Delta m^2_{\mbox{\rm{\scriptsize LSND}}}\sim{\cal O}(1)$eV$^2$.
If this anomaly as well as the atmospheric and solar neutrino data
are to be interpreted as evidence of neutrino oscillations then
we would need at least four flavors of neutrinos, since
the mass squared differences
$\Delta m^2_{\mbox{\rm\footnotesize atm}}$,
$\Delta m^2_\odot$ and
$\Delta m^2_{\mbox{\rm{\scriptsize LSND}}}$
suggested by the atmospheric neutrino
anomaly, the solar neutrino deficit and the LSND data have
different orders of magnitudes.

\section{Neutrino oscillations with three flavors}

\begin{figure}
\begin{center}
\vglue -1.0cm \hglue -0.5cm
\epsfig{file=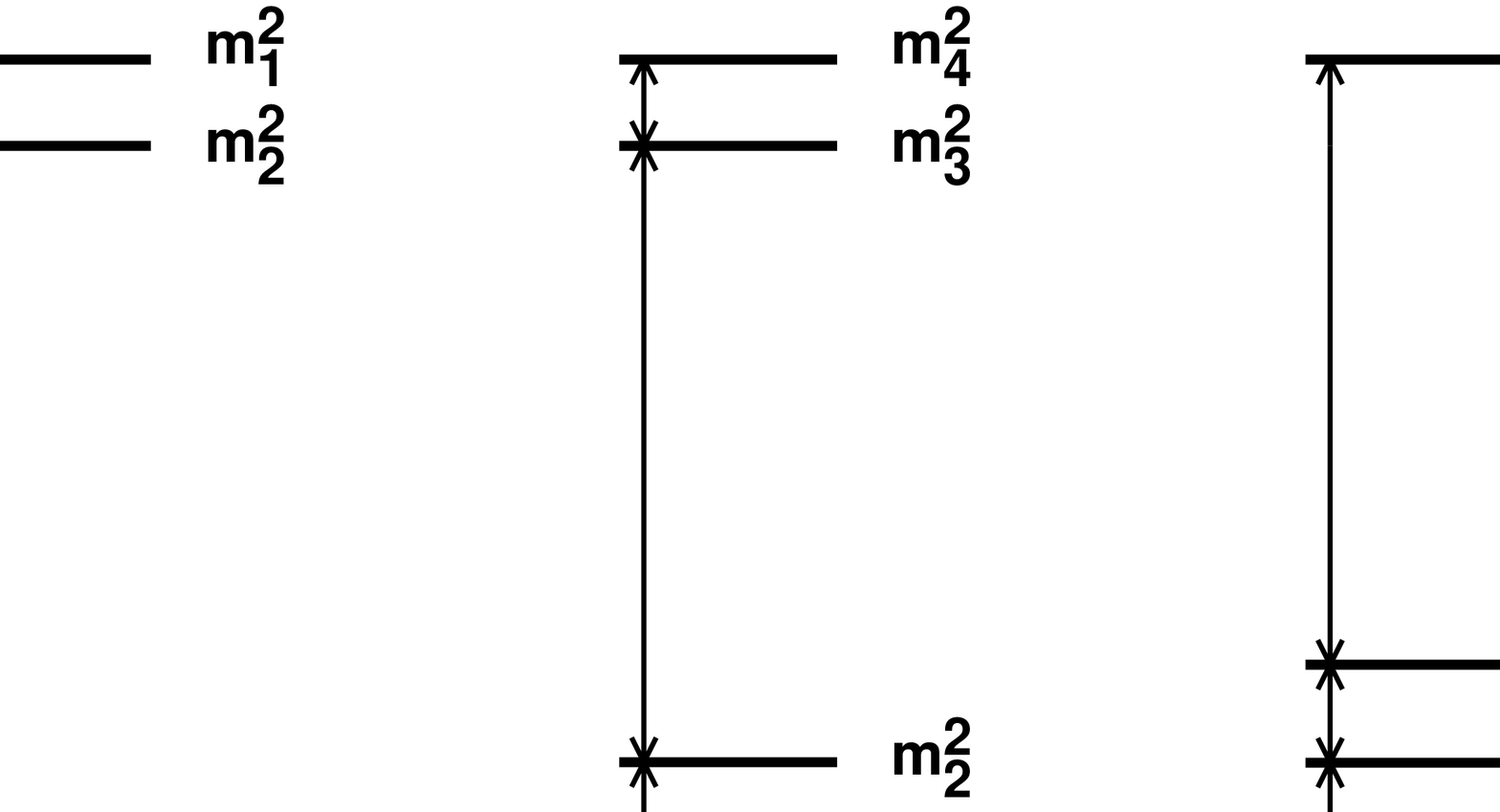,width=4cm}
\vglue 0.9cm
\caption{Mass patterns of three and four neutrino schemes.
(a) $N_\nu=3$ with $\Delta m^2_{32}>0$.
(b) $N_\nu=3$ with $\Delta m^2_{32}<0$.
(c) $N_\nu=4$ (2+2)-scheme, where either
($|\Delta m^2_{21}|=\Delta m^2_\odot$,
$|\Delta m^2_{43}|=\Delta m^2_{\mbox{\rm\footnotesize atm}}$)
or
($|\Delta m^2_{43}|=\Delta m^2_\odot$,
$|\Delta m^2_{21}|=\Delta m^2_{\mbox{\rm\footnotesize atm}}$).
The former is assumed in the subsection \ref{subsec:22}.
(d) and (e) are $N_\nu=4$ (3+1)-scheme, where
$|\Delta m^2_{41}|=\Delta m^2_{\mbox{\rm{\scriptsize LSND}}}$ and
either
($|\Delta m^2_{21}|=\Delta m^2_\odot$,
$|\Delta m^2_{32}|=\Delta m^2_{\mbox{\rm\footnotesize atm}}$)
or
($|\Delta m^2_{32}|=\Delta m^2_\odot$,
$|\Delta m^2_{21}|=\Delta m^2_{\mbox{\rm\footnotesize atm}}$) is satisfied.}
\label{fig:pattern}
\end{center}
\end{figure}

The flavor eigenstates are related to the mass eigenstates
by the $3\times3$
Pontecorvo-Maki-Nakagawa-Sakata
\cite{Pontecorvo:1957cp,Maki:1962mu}
(PMNS) mixing matrix:
\begin{eqnarray}
\left(
\begin{array}{ccc}
U_{e1} & U_{e2} &  U_{e3}\\
U_{\mu 1} & U_{\mu 2} & U_{\mu 3} \\
U_{\tau 1} & U_{\tau 2} & U_{\tau 3}
\end{array}\right)=
\left(
\begin{array}{ccc}
c_{12}c_{13} & s_{12}c_{13} &  s_{13}e^{-i\delta}\\
-s_{12}c_{23}-c_{12}s_{23}s_{13}e^{i\delta} &
c_{12}c_{23}-s_{12}s_{23}s_{13}e^{i\delta} & s_{23}c_{13}\\
s_{12}s_{23}-c_{12}c_{23}s_{13}e^{i\delta} &
-c_{12}s_{23}-s_{12}c_{23}s_{13}e^{i\delta} & c_{23}c_{13}\\
\end{array}
\right),\nonumber
\end{eqnarray}
and without loss of generality I assume $|\Delta
m_{21}^2|<|\Delta m_{32}^2|<|\Delta m_{31}^2|$
where $\Delta m^2_{ij}\equiv m^2_i-m^2_j$,
$m_j^2 (j=1,2,3)$ are the mass squared for the mass
eigenstates, and the different two mass patterns
(Fig. \ref{fig:pattern}(a)) and (Fig. \ref{fig:pattern}(b))
are distinguished by the sign of $\Delta m_{32}^2$.
With three flavors of neutrinos, it is
impossible to account for the solar neutrino deficit, the atmospheric
neutrino anomaly and LSND,
so I have to give up an effort to explain LSND and I have to
take $\Delta m^2_{21}=\Delta m^2_\odot$ and
$\Delta m^2_{32}=\Delta m^2_{\mbox{\rm{\scriptsize atm}}}$.
Under the present assumption it follows
$\Delta m^2_{\rm atm}=\Delta m^2_{32}$
$\gg$$\Delta m^2_{21}=\Delta m^2_\odot$
and I have a large hierarchy between
$\Delta m^2_{21}$ and $\Delta m^2_{32}$.
If $|\Delta m^2_\odot L/4E|\ll 1$ then from a hierarchical
condition I have the oscillation probability
\begin{eqnarray}
P(\bar{\nu}_e \rightarrow \bar{\nu}_e) 
= 1-\sin^22\theta_{13}\sin^2(\Delta m_{32}^2 L/4E),\nonumber
\end{eqnarray}
so if $\Delta m^2_{\rm atm} > 1.5\times10^{-3}$eV$^2$
then the CHOOZ reactor data \cite{Apollonio:1999ae}
force us to have
either $\theta_{13}\simeq 0$ or $\theta_{13}\simeq \pi/2$.
To account for the solar neutrino deficit
$|s_{13}|$ cannot be too large, so
by combining the atmospheric neutrino data
it follows that $|\theta_{13}|\ll 1$ and the 
PMNS mixing matrix $U$
becomes \cite{Yasuda:1999uv,Fogli:2000ak,Gonzalez-Garcia:2001sq}
\begin{eqnarray}
U_{PMNS} \simeq
\left(
\begin{array}{ccc}
c_\odot & s_\odot &  \epsilon\\
-s_\odot/\sqrt{2} &
c_\odot/\sqrt{2} & 1/\sqrt{2}\\
s_\odot/\sqrt{2} &
-c_\odot/\sqrt{2} & 1/\sqrt{2}\\
\end{array}
\right),
\label{eqn:mns3}
\end{eqnarray}
which indicates that the solar neutrino problem is explained
by oscillations half of which is
$\nu_e\rightarrow\nu_\mu$ and the other of which is
$\nu_e\rightarrow\nu_\tau$, and that
the atmospheric neutrino anomaly is accounted for
by oscillations of almost 100\% $\nu_\mu\rightarrow\nu_\tau$
($|\epsilon|\equiv|\theta_{13}|\ll 1$).  In (\ref{eqn:mns3})
$\theta_\odot=\theta_\odot$(LMA) or $\theta_\odot$(LOW)
at 90\%CL.

\section{Neutrino oscillations with four flavors}

In the case of four neutrino schemes there are two distinct types of
mass patterns.  One is the so-called (2+2)-scheme
(Fig. \ref{fig:pattern}(c)) and the other is the (3+1)-scheme
(Fig. \ref{fig:pattern}(d) or (e)).  Depending on the type of the two
schemes, phenomenology is different.

\subsection{(2+2)-scheme}\label{subsec:22}

The combined analysis of
the atmospheric and solar neutrino data with and without the
SNO data \cite{Ahmad:2001an} was done by Ref.~\cite{Gonzalez-Garcia:2001zi}
and Ref.~\cite{Gonzalez-Garcia:2001uy}
in the (2+2)-scheme.
Without the SNO data the best fit point is close to
pure $\nu_e\leftrightarrow\nu_s$ with the small mixing
angle (SMA) MSW solution for $\nu_\odot$
and pure $\nu_\mu\leftrightarrow\nu_\tau$ for $\nu_{\rm atm}$
with maximal mixing:
\begin{eqnarray}
U_{PMNS}
=\left(
\begin{array}{cccc}
U_{e1} & U_{e2} &  U_{e3} &  U_{e4}\\
U_{\mu 1} & U_{\mu 2} & U_{\mu 3} & U_{\mu 4}\\
U_{\tau 1} & U_{\tau 2} & U_{\tau 3} & U_{\tau 4}\\
U_{s1} & U_{s2} &  U_{s3} &  U_{s4}
\end{array}\right)
\simeq\left(
\begin{array}{cccc}
c_\odot & s_\odot & \epsilon & \epsilon\\
\epsilon & \epsilon&1/\sqrt{2} & 1/\sqrt{2} \\
\epsilon & \epsilon&-1/\sqrt{2} & 1/\sqrt{2} \\
-s_\odot & c_\odot & \epsilon & \epsilon\\
\end{array}\right),
\label{eqn:sma}
\end{eqnarray}
where $|\epsilon|\ll 1$,
$\sin^22\theta_\odot=\sin^22\theta_\odot({\rm SMA})\sim 10^{-3}$.
With the SNO result the best fit solution is described by
\begin{eqnarray}
U_{PMNS}
=\left(
\begin{array}{cccc}
U_{e1} & U_{e2} &  U_{e3} &  U_{e4}\\
U_{\mu 1} & U_{\mu 2} & U_{\mu 3} & U_{\mu 4}\\
U_{\tau 1} & U_{\tau 2} & U_{\tau 3} & U_{\tau 4}\\
U_{s1} & U_{s2} &  U_{s3} &  U_{s4}
\end{array}\right)
\simeq\left(
\begin{array}{cccc}
c_\odot & s_\odot & \epsilon & \epsilon\\
U_{\mu 1} & U_{\mu 2} & U_{\mu 3} & U_{\mu 4}\\
U_{\tau 1} & U_{\tau 2} & U_{\tau 3} & U_{\tau 4}\\
U_{s1} & U_{s2} &  U_{s3} &  U_{s4}
\end{array}\right),
\label{eqn:lma}
\end{eqnarray}
where $\theta_\odot=\theta_\odot({\rm LMA})$
and $|U_{s1}|^2+ |U_{s2}|^2\simeq 0.2$.
In the solution (\ref{eqn:lma}) the ratio of active and
sterile oscillations in $\nu_\odot$ is 80\% and 20\%, respectively,
while that of active and
sterile oscillations in $\nu_{\rm atm}$ is approximately
30\% and 70\% at the maximum of
$\sin^2(\Delta m^2_{\rm atm}L/4E)$, respectively.
The reason that dominant sterile oscillation
gives a good fit
to the atmospheric neutrino data is because the disappearance
probability can contain a constant term $B$ which
serves as an extra free parameter \cite{Yasuda:2000de} \footnote{
In Ref.~\cite{Yasuda:2000de} the atmospheric data 
was analyzed in the (2+2)-scheme with a CP phase $\delta_1$
which was ignored in
Refs.~\cite{Gonzalez-Garcia:2001zi,Gonzalez-Garcia:2001uy},
but the results with $\delta_1=\pi/4,\pi/2$ in Ref.~\cite{Yasuda:2000de}
are more or less the same as those with $\delta_1=0$.}:
$1-P(\nu_\mu\rightarrow\nu_\mu)=
A\sin^2(\Delta m^2_{\rm atm}L/4E)
+B\sin^2(\Delta m^2_{\mbox{\rm{\scriptsize LSND}}}L/4E)
\rightarrow A\sin(\Delta m^2_{\rm atm}L/4E)+B/2$,
where I have averaged over rapid oscillations.
The goodness of fit for the mixing (\ref{eqn:lma})
is 67\% ($\chi^2$=73.8 for 80 degrees of freedom), which is quite
good.  On the other hand, the mixing (\ref{eqn:sma}) with
the SMA MSW solution still gives a good fit even with the SNO result
(the goodness of fit is 62\%, or $\chi^2$=75.6 for 80 degrees of
freedom), so that pure sterile oscillations in $\nu_\odot$
plus pure active oscillations in $\nu_{\rm atm}$ is still an
acceptable solution in the four flavor framework
despite the SNO data.  To exclude the (2+2)-scheme,
therefore, one needs to improve much more statistics and systematics
both in the atmospheric and solar neutrino data.

\subsection{(3+1)-scheme}

It has been shown in Refs.~\cite{Okada:1997kw,Bilenkii:1998rw}
using the older LSND result that
the (3+1)-scheme is inconsistent with the Bugey
reactor data \cite{bugey} and the CDHSW disappearance experiment \cite{cdhsw}
of $\nu_\mu$.  However, in the final result \cite{Mills:2001tq}
the allowed region has shifted
to the lower value of $\sin^22\theta$ and it was shown \cite{bklw}
that there are four isolated regions
$\Delta m^2_{\mbox{\rm{\scriptsize LSND}}}\simeq$0.3, 0.9, 1.7, 6.0 eV$^2$
which satisfy all the constraints of Bugey, CDHSW and
the LSND data (99\%CL).
The case of $\Delta m^2_{\mbox{\rm{\scriptsize LSND}}}$=0.3 eV$^2$
is excluded by the SuperKamiokande atmospheric neutrino
data.  For the other three values of
$\Delta m^2_{\mbox{\rm{\scriptsize LSND}}}$,
the best fit solution looks like
\begin{eqnarray}
U_{PMNS}
=\left(
\begin{array}{cccc}
U_{e1} & U_{e2} &  U_{e3} &  U_{e4}\\
U_{\mu 1} & U_{\mu 2} & U_{\mu 3} & U_{\mu 4}\\
U_{\tau 1} & U_{\tau 2} & U_{\tau 3} & U_{\tau 4}\\
U_{s1} & U_{s2} &  U_{s3} &  U_{s4}
\end{array}\right)
\simeq\left(
\begin{array}{cccc}
c_\odot &  {\ }s_\odot& {\ }0 & {\ }\epsilon\\
-s_\odot/\sqrt{2}& {\ }c_\odot/\sqrt{2} &  {\ }{1 \over \sqrt{2}} &{\ }\delta\\
s_\odot/\sqrt{2} & {\ }-c_\odot/\sqrt{2} &  {\ }{1 \over \sqrt{2}} & {\ }0\\
-{\epsilon \over \sqrt{2}}-{\delta \over 2}&
{\ }-{\epsilon \over \sqrt{2}}+{\delta \over 2} &   {\ }0 & {\ }1
\end{array}\right),
\label{eqn:31}
\end{eqnarray}
where $|\epsilon|,|\delta|\ll 1$,
$\theta_\odot=\theta_\odot$(LMA).
Since the off diagonal elements are small,
the solution (\ref{eqn:31}) gives almost the same phenomenology
as that of the three flavor scenario, so that
it is difficult to exclude this scheme by atmospheric
or solar neutrino experiments.
Hence it will remain a viable scheme until the LSND data are disproved
by the new experiment MiniBooNE \cite{miniboone}.

\section{Long baseline experiments}

The only long baseline experiment which is already running is K2K
\cite{Ahn:2001cq}, and it is expected to give us more precise value
of $\Delta m^2_{\rm atm}$ than the atmospheric neutrino
observations.  The next future long baseline
experiments such as MINOS \cite{minos}, OPERA \cite{opera} and
JHF \cite{Itow:2001ee} will determine more precisely the values of $\Delta
m^2_{\rm atm}$ and $\sin^22\theta_{\rm atm}$, and possibly measure the
value of $\sin^22\theta_{13}$ by looking at appearance of $\nu_e$
in $\nu_\mu$ beam:
\begin{eqnarray}
P(\nu_\mu (\bar{\nu}_\mu )\rightarrow \nu_e(\bar{\nu}_e))
=s^2_{23}\sin^22\theta_{13}
\left({\Delta E_{32} \over \Delta E^{M(\pm)}_{32}}\right)^2
\sin^2\left(\Delta E^{M(\pm)}_{32} L \over 2\right),\nonumber
\end{eqnarray}
where
\begin{eqnarray}
\Delta E^{M(\pm)}_{32}\equiv
\sqrt{\left(\Delta E_{32}\cos2\theta_{13}\pm \sqrt{2}G_F N_e\right)^2
+\left(\Delta E_{32}\sin2\theta_{13}\right)^2},\nonumber
\label{eqn:pem}
\end{eqnarray}
$\Delta E_{32}\equiv\Delta m^2_{32}/2E$, $N_e$ stands for the
electron density of matter, and
$-\sqrt{2}G_F N_e$, $+\sqrt{2}G_F N_e$
stands for the matter effect for neutrinos and for anti-neutrinos,
respectively.
In further future the second stage JHF experiment with 4MW and/or
neutrino factory \cite{Geer:1998iz} may be able to measure the value
of sign$(\Delta m^2_{32})$, which is crucial to determine the mass
pattern of three neutrino schemes, and the value of the CP phase
from the difference in the appearance probabilities for
neutrinos and for anti-neutrinos.
Currently there are a lot of issues which are subjects of active research,
such as treatment of the uncertainty of the matter effect
in measurements of CP violation, each advantage of high energy or
low energy option, correlations of errors, etc.
Details of recent developments in long baseline experiments and
neutrino factories can be found on the web page of nufact'01
\cite{nufact01}.

\section*{Acknowledgments}
I would like to thank Prof. Kyungsik Kang and other organizers
for invitation and hospitality
during the workshop.  This
research was supported in part by a Grant-in-Aid for Scientific
Research of the Ministry of Education, Science and Culture,
\#12047222, \#13640295.

\end{document}